\documentclass[prd,10pt,twocolumn,nofootinbib,preprintnumbers]{revtex4}
\usepackage{amsmath,amssymb,slashed,mathtools,slashed}
\usepackage{graphicx,multirow}
\usepackage{units}
\usepackage[hyperfootnotes=false,colorlinks,citecolor=blue]{hyperref}
\usepackage{mathtools}
\usepackage{pifont}
\usepackage{amsmath,amssymb}
\usepackage{epsfig}
\usepackage{url}
\usepackage{subfigure}
\usepackage{graphicx}
\usepackage{grffile}
\usepackage[usenames,dvipsnames]{color}
\usepackage{slashed}
\usepackage{array}
\usepackage{multirow}
\usepackage{braket}
\usepackage[colorlinks,citecolor=blue]{hyperref}
\usepackage{color}
\usepackage{cancel}
\usepackage{gensymb}

\def\a {\alpha}

\def\l {\lambda}
\def\L {\Lambda}

\def\bar {\overline}

\def\be {\begin{equation}}
\def\ee {\end{equation}}
\def\beq {\begin{equation}}
\def\eeq {\end{equation}}
\def\bea {\begin{eqnarray}}
\def\eea {\end{eqnarray}}

\newcommand{\besub}{\begin{subequations}}
\newcommand{\eesub}{\end{subequations}}

\def\beq{\begin{equation}}
\def\eeq{\end{equation}}
\def\barr{\begin{array}}
\def\earr{\end{array}}

\begin{document}
\title{Limiting multiple hyperchargeless scalar triplets using electroweak phase transition and other constraints}

\author{Nabarun Chakrabarty}
\email{nabarunc@iitk.ac.in}
\affiliation{Department of Physics, Indian Institute of Technology Kanpur, Kanpur, Uttar Pradesh-208016, India}

\begin{abstract} 

While an additional scalar multiplet over and above the Standard Model can lead to a strong electroweak phase transition, depending on its quantum numbers, it also potentially confronts crucial constraints from theory and experiments. Should there exist more than one copy of such a multiplet, it is possible to predict that number from the requirements of a strong electroweak phase transition and agreement with the latest constraints. We aim to probe this specific issue in this study in the context of
$N$ degenerate scalar triplets governed by a global O$(N)$ symmetry. We fold in important constraints from $h \to \gamma\gamma$ signal strength, dark matter direct detection and Landau pole behaviour. A combined analysis reveals $N \gtrsim 70$ for a strong phase transition and consistency with the constraints. We also look into possible gravitational wave signals in the parameter regions of interest.
 
\end{abstract} 
\maketitle

\section{Introduction} 

While the Higgs discovery~\cite{ATLAS:2012yve,CMS:2012qbp} at the Large Hadron Collider (LHC) finds the last missing piece in the Standard Model (SM), 
the dynamics governing particle interactions is still not fully understood. 
It has been shown that though the SM Higgs can generate a first order electroweak phase transition (EWPhT), it is not \emph{strong}~\cite{KAJANTIE1996189,Bochkarev:1987wf,Michela2014}, as it must be in order to successfully account for the observed imbalance between baryons and anti-baryons in the universe~\cite{KUZMIN198536,SHAPOSHNIKOV1987757,
Shaposhnikov1986465}, i.e., $\eta = (n_b - n_{\bar{b}})/n_\gamma$ = 5.8-6.6 $\times 10^{-10}$. Nor can the SM Higgs predict a stable electroweak vacuum till the Planck scale for all values of the $t$-mass~\cite{Buttazzo:2013uya,Degrassi:2012ry,Tang:2013bz,Ellis:2009tp,Elias-Miro:2011sqh}. Such shortcomings seriously vouch for the presence of an additional scalar sector. In fact, an extended scalar sector can also be enticing from the perspective of dark matter. That is, an electrically neutral scalar emerging from such a scalar sector can explain the observed dark matter relic abundance~\cite{Planck:2018vyg}.  

The advent of gravitional wave (GW) astronomy has made it possible to probe GW spectra associated with a strong first order phase transition~\cite{Kamionkowski:1993fg,Witten:1984rs,Hogan:1986qda,Grojean:2006bp}. A first order phase transition originates from the bubble nucleation of the true vacuum. When a bubble expands, the broken phase (or true vacuum) it contains within extends to the unbroken phase (or false vacuum) outside. Such an expansion causes the bubbles to collide and thereby emit GWs~\cite{LINDE1983421,Witten:1984rs,Hogan:1986qda,Caprini:2019egz,Gould:2019qek,Weir:2017wfa,Hindmarsh:2017gnf,Guo:2020grp,Hindmarsh:2015qjv}. Such GWs can potentially be detected in the proposed space-based interferometers such as LIGO~\cite{PhysRevLett.116.241103}, LISA~\cite{LISA:2017pwj}, DECIGO~\cite{Yagi:2011wg} etc. In all, a given scalar extension can therefore be probed through its GW imprints.

While the quantum numbers of an extended Higgs sector can have a multitude of possibilities, the ones with non-trivial SM charges can have more attractive features through their interactions with the SM gauge bosons. Such multiplets confront the constraints from the $h \to \gamma \gamma$ signal strength. Further, they can also induce sizeable radiative effects to measurable amplitudes having gauge bosons in the final states. A foremost example is a radiatively corrected $h W W$ or $h Z Z$ vertex with the latter's strength directly connected to Higgs production at the proposed International Linear Collider~\cite{ILC:2007oiw,ILC:2007bjz,Phinney:2007gp,ILC:2007vrf,Behnke:2013lya}. In fact, there could be more copies of such a scalar multiplet. While 
a strong first order phase transition would favour a larger number of multiplets, other constraints such as the absence of a Landau pole below a certain cutoff scale, the $h \to \gamma \gamma$ constraint and the dark matter direct detection bound would tend to limit that number. And this opens the possibility of an interesting interplay. In this study, we aim to determine the minimum number of a scalar multiplet that catalyses strong EWPhT and is also simultaneously consistent with miscellaneous restrictions from both theory and experiments. We also explore if the number for the multiplet so obtained can lead to observable GW signals at the proposed detectors.

We now try to zero in on the choice of quantum numbers for the scalar multiplet. The most minimal one confronting $h \to \gamma\gamma$ is a (\textbf{1,1},$Q$) scalar under the SM gauge group when $Q \neq 0$. This choice is interesting and has been explored in~\cite{Ahriche:2018rao}. However, such a scalar cannot be DM candidate owing to the non-zero electric charge. The next minimal multiplet in terms of degrees of freedom is (\textbf{1,3},0) that comprises three real scalars, or, a complex scalar and a real scalar in the standard sense~\cite{Araki:2011hm,Bandyopadhyay:2013lca,Bandyopadhyay:2014vma,Bell:2020hnr}. Such ae introduce $N$ copies of a ($1,3,0$) triplet governed by a global O$(N)$ symmetry. That is, a given triplet transforms as $\Delta_i \to O_{ij} \Delta_j$, where $i,j=1,...N$ and $O_{ij}$ is an element of the orthogonal matrix. A particular triplet $\Delta_i$ can be expressed as
\bea
\Delta_i &=&
\begin{pmatrix}
T_{0i}/\sqrt{2} & T_i^+ \\
T_i^- & -T_{0i}/\sqrt{2}
\end{pmatrix}.
\eea
The lagrangian consistent with the gauge and $O(N)$ symmetry is
\bea
\mathcal{L} &=& \mathcal{L}_{\text{SM}} + \frac{1}{2}\sum^N_{i=1} 
\text{Tr}\Big[(D_\mu \Delta_i) (D^\mu \Delta_i) \Big] + \frac{1}{2}\mu_\Delta^2 \sum^N_{i=1}
\text{Tr}[\Delta_i \Delta_i] \nonumber \\
&&
 + \l_{H \Delta} \sum^N_{i=1} H^\dagger H ~\text{Tr}[\Delta_i \Delta_i]
 + \l_\Delta \Bigg( \sum^N_{i=1}
\text{Tr}[\Delta_i \Delta_i] \Bigg)^2.
\eea
It is pointed out that the trace is over the $SU(2)_L$ indices. All parameters in the scalar potential are chosen real to avoid CP-violation. We take $\mu_\Delta^2 > 0$ in this study which ensures that $\Delta_i$ does not recieve a vacuum expectation value. This forbids mixing among the triplets and leads to an $N$-fold degenerate mass spectrum. The common mass for $T_{0i}$ is given by
\besub
\bea
m^2_{T_{0i}} &=& \mu^2_\Delta + \frac{\l_{H\Delta}}{2} v^2
~~\text{for}~i=1,...N; \\
& \equiv & m^2_{T_0}.
\eea
\eesub
The common mass for $T^+_{i}$, say $m_{T^+}$, is slightly deviated from $m_{T_0}$ on account of radiative effects~\cite{Sher:1995tc,Cirelli:2005uq}. That is
\bea
m_{T^+} - m_{T_0} &=& \frac{\a m_{T_0}}{4\pi}\bigg[ 
f\Big( \frac{M_W}{m_{T_0}} \Big) - c^2_W f\Big( \frac{M_Z}{m_{T_0}} \Big)   \bigg]. 
\eea
Here, $\a$ is the fine-structure constant and $f(x) = -\frac{x}{4} [2 x^3 \text{ln}(x) + (x^2 - 4)^{3/2}\text{ln}\big( \frac{x^2 - 2 -x\sqrt{x^2 - 4}}{2} \big)]$. One derives this splitting to be $\simeq$ 166 MeV for 100 GeV $< m_{T_0} <$ 2 TeV. Eliminating $\mu_\Delta$ in favour of $m_{T_0}$, we use $\{m_{T_0},\l_{H\Delta},\l_{\Delta},N \}$ to describe the triplet sector. That the number of Lagrangian parameters for $N$ triplets remains the same as the $N$=1 case is an artefact of the O($N$) invariance.

This study is organised as follows. We discuss the theoretical constraints on this setup in section \ref{theory}. Discussions on oblique parameters and the diphoton signal strength are to be found in sections \ref{oblique} and \ref{diphoton}. We elaborate the dark matter constraint, especially direct detection in section \ref{dm}.
The dynamics of a SFOEWPhT and the generation of GW in context of the present framework is described in section \ref{PT}. The same section
also contains an analysis combining the SFOEWPhT and the aforesaid constraints. We finally conclude in section \ref{conclu}.

\section{Theory constraints}\label{theory}

\subsection{Perturbativity} We first discuss possible bounds on the framework from perturbativity, albeit in a somewhat heuristic fashion. We do a naive power counting analysis of $\Delta_i~\Delta_j \to \Delta_k~\Delta_l$ using the first few terms of the perturbation series. It shows that the successive terms get smaller in magnitude for $\frac{N\l_\Delta}{16\pi^2} \lesssim 1$. In fact, a similar exercise in case of $\Delta_i~\Delta_j \to \phi \phi$ also leads to the same condition. Thus, the bound on $\l_\Delta$ is
\bea
\l_\Delta \lesssim \frac{16 \pi^2}{N}.
\eea
However, when it comes to $\phi \phi \to \phi \phi$, we prima facie end up with two conditions, i.e., $\frac{N \l^2_{H\Delta}}{16 \pi^2} < \l_H$ and $\frac{N \l_{H\Delta}}{16 \pi^2} < 1$. That is,
\bea
\l_{H\Delta} \lesssim 4\pi \sqrt{\frac{\l_H}{N}}~\text{and}
~\l_{H\Delta} \lesssim \frac{16\pi^2 }{N}. 
\eea
We obey the more stringent of the two bound in our analysis. One finds that $\l_{H\Delta} \lesssim 4\pi \sqrt{\frac{\l_H}{N}}$ is more constraining for $N \lesssim 1200$.

\subsection{Location of Laudau poles}

An $O(N)$-multiplet can lead to interesting predictions on the location of Landau poles of the theory. Since the present scenario involves only two additional scalar quartic couplings $\l_{HT}$ and $\l_T$ in addition to the SM-like Higgs self coupling $\l_H$, it is straightforward to understand such dynamics in terms of the one-loop beta functions. We quote below the one-loop beta functions for this framework that are dependent on $N$.
\besub
\bea
16 \pi^2 \beta_{\l_H} &=& 24 \l^2_H + \frac{3}{2} N \l^2_{HT}, \\
16 \pi^2 \beta_{\l_{HT}} &=& 12 \l_H \l_{HT} + 4 \l^2_{HT} \nonumber \\
&&
 + \bigg( \frac{2 + 3N}{3} \bigg) \l_{T}\l_{HT}, \\
16 \pi^2 \beta_{\l_{T}} &=& 12 \l^2_{HT} + \bigg( \frac{8 + 3N}{3} \bigg) \l^2_{T}, \\
16 \pi^2 \beta_{g_2} &=& \bigg(-\frac{19}{6} + \frac{N}{3} \bigg) g_2^3.
\eea
\eesub
The beta functions for the rest of the parameters coincide with the corresponding SM expressions and hence are not given here for brevity. One reckons that the Landau pole shifts to lower energy scales with increasing $N$ for fixed values of the couplings at the input scale. For illustration, the Landau pole for $g_2$ at one-loop can be analytically determined to be $\L^\prime = M_t~\text{exp}\bigg[ \frac{8 \pi^2}{\big(-\frac{19}{6} + \frac{N}{3}\big)\big(g_{2}(M_t)\big)^2} \bigg]$. We then find that $\L^\prime \gtrsim$ 1 TeV (10 TeV) requires $N \lesssim 330~(148)$. The choice of $\L$ is driven by the expectation that some hitherto unknown additional dynamics takes over beyond this scale. In this work, we ensure the absence of Landau poles below $\L$ by demanding the more stringent condition of perturbativity, i.e., the magnitudes of the quartic (gauge) couplings remain smaller than $4\pi$ ($\sqrt{4\pi}$) at all intermediate scales up to $\L$ under renormalisation group evolution. This is a more stringent restriction. For example, demanding a perturbative $g_2$ up to 1 TeV (10 TeV) necessitates $N \lesssim 320~(144)$.

\section{Oblique parameters}\label{oblique}

A function that appears while calculating the contribution of a BSM scalar sector to the oblique paramater $T$ is $F(x,y)=\frac{x+y}{2}-\frac{x y}{x - y}\text{ln}\Big( \frac{x}{y} \Big)$. We then quote below the contribution of the $O(N)$-symmetric $Y=0$ triplet sector to $T$.
\besub
\bea
\Delta T &=& \frac{N}{16 \pi s^2_W M^2_W}F(m^2_{T^+},m^2_{T_0}), \\
&\simeq& \frac{N(m_{T^+} - m_{T^0})^2}{24 \pi s^2_W M^2_W} \simeq 2.458 \times 10^{-7}~N.
\eea
\eesub
It is immediately seen that an $N$ consistent with the perturbativity constraint, for $g_2$ for instance, $\Delta T$ is well-within the latest PDG bound~\cite{Bell:2020hnr}. There thus is no constraint from the $T$-parameter in this framework. 

\section{$h \to \gamma \gamma$ signal strength}\label{diphoton}

The $O(N)$-symmetric scalar triplet sector does not modify the tree level couplings of $h$ with other SM particles on account of zero mixing between $\phi$ and $\Delta_i$. However, it 
does modify the loop-induced $h \to \gamma \gamma$ amplitude through additional loops
involving the charged scalars $T_i^\pm$. The O$(N)$ symmetry ensures that the $T^+_i-T^-_i-h$ coupling is $-\l_{H\Delta}v$ for all $i=1,...,N$. The $h \to \gamma \gamma$ amplitude induced by the 
$O(N)$-symmetric sector therefore reads
\bea
\mathcal{M}^{O(N)}_{h \to \gamma \gamma} &=& N \frac{\l_{H \Delta} v^2}{2 m^2_{T^+}} A_0\bigg(\frac{m^2_h}{4 m^2_{T^+}}\bigg).
\label{htogaga_np}
\eea
Here $A_0(x)$ is the amplitude for the spin-0 particles in the loop~\cite{Djouadi:2005gi,Djouadi:2005gj} and is expressed as
\bea
A_0(x) &=& -\frac{1}{x^2}\big(x - f(x)\big),~\text{with}
\eea
\bea
f(x) &=& \text{arcsin}^2(\sqrt{x}); ~x \leq 1 
\nonumber \\
&&
= -\frac{1}{4}\Bigg[\text{log}\frac{1+\sqrt{1 - x^{-1}}}{1-\sqrt{1 - x^{-1}}} -i\pi\Bigg]^2; ~x > 1.
\eea
The $O(N)$ symmetry expectedly makes the contribution of the triplets to scale with $N$. The amplitude and the decay width for this model is thus given by
\bea
\mathcal{M}_{h \to \gamma \gamma} &=& 
\mathcal{M}^{\text{SM}}_{h \to \gamma \gamma} +
\mathcal{M}^{O(N)}_{h \to \gamma \gamma}, 
\\
\Gamma_{h \to \gamma \gamma} &=& \frac{G_F \a_{\text{em}}^2 m_h^3}{128 \sqrt{2} \pi^3} |\mathcal{M}_{h \to \gamma \gamma}|^2.\eea
where $G_F$ and $\a_{\text{em}}$ denote respectively the Fermi constant and the QED fine-structure constant.

Since the $pp\to h$ production rate remains unchanged in this model and the modification to the total width is negligible, the signal strength becomes
\bea
\mu_{\gamma\gamma}
& \simeq & \frac{\Gamma^{\text{SM}}_{h\to\gamma\gamma}}{\Gamma_{h\to\gamma\gamma}}.
\eea 
The latest 13 TeV results on the diphoton signal strength from
the LHC read $\mu_{\gamma\gamma} = 0.99^{+0.14}_{-0.14}$ (ATLAS~\cite{ATLAS:2018hxb}) and
$\mu_{\gamma\gamma} = 1.18^{+0.17}_{-0.14}$ (CMS~\cite{CMS:2018piu}). We obtain $\mu_{\gamma\gamma} = 0.99 \pm 0.14$ upon combining the individual central values and uncertainties and impose this constraint at 2$\sigma$.

\section{Dark matter}\label{dm}

The degenerate neutral scalars $T_{0i}$ are rendered cosmologically stable by virtue of the $O(N)$ symmetry and hence they become potential DM candidates. In fact, the $N=1$ (a single $Y=0$ triplet) case is widely studied where a $\mathbb{Z}_2$-like discrete symmetry is required to stabilise the neutral scalar. The tiny difference between the masses of $T^+$ and $T_0$ compulsorily triggers coannihilation processes of the type $T^+ T_0 \to W^+ \gamma, W^+ Z, W^+ h$. Overall, the single triplet case is seen to yield the observed relic sensity, i.e., $\Omega^{\text{obs}}_{\text{DM}} h^2 = 0.1200 \pm 0.001$~\cite{Planck:2018vyg}, only for $m_{T_0} \gtrsim 1.8$ TeV. 

In  the case of multiple DM candidates, the total relic density is always a sum of the individual relic densities. Given the individual $T_{0i}$ are mass-degenerate in this setup having identical interaction strengths, the total relic density simply becomes
\bea
\Omega_{\text{total}} h^2 = N~\Omega_{T_0} h^2,
\eea 
where $\Omega_{T_0} h^2$ is the value in case of $N$ = 1. In this study, we do not impose the prediction of the observed relic as a binding requirement and impose rather the relaxed condition $\Omega_{\text{total}} h^2 \leq 0.12$. That is, we keep alive the possibility that there are other DM sources in the universe
that account for the rest of the requisite relic. 

We discuss next our treatment of the direct detection bounds in this analysis. Let us assume DM entity, say $\chi$, predicts a relic density $\Omega_\chi h^2 \leq \Omega^{\text{obs}}_{\text{DM}} h^2$ and a direct detection cross section 
$\sigma^{\text{DD}}_{\chi}$. Then, the \emph{effective} direct detection cross section is computed as $\sigma^{\text{DD}}_{\chi,~\text{eff}} = \bigg(\frac{\Omega_\chi h^2}{\Omega^{\text{obs}}_{\text{DM}} h^2}\bigg)\sigma^{\text{DD}}_{\chi}$. The direct detection constraints are obeyed by stipulating that $\sigma^{\text{DD}}_{\chi,~\text{eff}}$ stays within the bound set by the experiments. The fraction in the prefactor accordingly scales down the theoretical cross section in case of underabundance. Applying this to the $O(N)$ scalars, one writes the effective direct detection cross section for a given $T_{0i}$ as
\bea
\sigma^{\text{DD}}_{i,~\text{eff}} = \bigg(\frac{\Omega_{T_0} h^2}{\Omega^{\text{obs}}_{\text{DM}} h^2}\bigg)\sigma^{\text{DD}}_i \simeq  \bigg(\frac{\Omega_{T_0} h^2}{0.12}\bigg)\sigma^{\text{DD}}_{T_0}. \label{DD}
\eea
We use $\sigma^{DD}_i = \sigma^{DD}_{T_0}$ in the last step in Eq.(\ref{DD}). We compute $\Omega_{T_0} h^2$ and 
$\sigma^{\text{DD}}_{T_0}$ using the publicly available tool \texttt{micrOMEGAs}~\cite{Belanger:2014vza} and demand that 
$\bigg(\frac{\Omega_{T_0} h^2}{0.12}\bigg)\sigma^{\text{DD}}_{T_0}$ stays below the bound from the XENON-1T experiment~\cite{XENON:2017vdw}.

On the other hand, indirect search experiments have put upper bounds on DM annihilation cross section to SM states such as $b\bar{b}$ and $\gamma\gamma$.
Since we allow for abundance, the quantity of interest is an effective indirect detection cross section defined as 
\bea
\sigma^{\text{ID}}_{i,~\text{eff}} = \bigg(\frac{\Omega_{T_0} h^2}{\Omega^{\text{obs}}_{\text{DM}} h^2}\bigg)^2 \sigma^{\text{ID}}_i \simeq  \bigg(\frac{\Omega_{T_0} h^2}{0.12}\bigg)^2\sigma^{\text{ID}}_{T_0}. \label{DD}
\eea 
One notes that the ratio 
$\bigg(\frac{\Omega_{T_0} h^2}{0.12}\bigg)$
gets squared for indirect detection since there now are two DM particles in the initial state.

\section{Finite temperature scalar potential and first order phase transitions}\label{PT}

We discuss in detail the $T \neq 0$ dynamics leading up to a first order phase transition for this setup. Investigations of phase transition in scenarios involving scalar triplets can be found in~\cite{Kazemi:2021bzj,Niemi:2018asa,Garcia-Pepin:2016hvs}. The background field $\phi$ in terms of which the scalar potential is expressed is chosen to be along the Higgs direction, i.e., one substitutes $H = \frac{1}{\sqrt{2}}\begin{pmatrix}
0 \\
 \phi
\end{pmatrix}$. Note that $\phi$ is distinct from the physical Higgs boson $h$. The tree level scalar potential as a function of the background field is given by
\bea
V_0(\phi) &=& -\frac{\mu^2_H}{2} \phi^2 + \frac{\l_H}{4} \phi^4.
\eea
Next, we quote the one-loop Coleman-Weinberg correction to the scalar potential following [Quiros].
\bea
\Delta V_{\rm CW}(\phi) &=& \frac{1}{64 \pi^2} \sum_a n_a \bigg[ m^4_a(\phi) \Big( \text{log} \frac{m_i^2(\phi)}{m_a^2(v)} - \frac{3}{2}\Big) \nonumber \\
&&
 + 2 m^2_a(v)m^2_a(\phi)\bigg],
\eea
where $n_a$ denotes the degrees of freedom of the $a$th particle.
This particular form for $V_{\rm CW}(\phi)$ ensures that the electroweak vacuum $v$ and the Higgs mass $M_h$ do not change upon the inclusion of one-loop corrections. The field-dependent squared masses $m^2_a(\phi)$ for this framework are given below.

\besub
\bea
m^2_W(\phi) &=& \frac{1}{4} g_2^2 \phi^2,~m^2_Z(\phi) = \frac{1}{4} (g_1^2 + g_2^2) \phi^2, \\
m^2_t(\phi) &=& \frac{1}{2} y^2_t \phi^2,~m^2_h(\phi) = 3 \l_H \phi^2 - \mu^2_H, \\
m^2_{T^0}(\phi) &=& m^2_{T^+}(\phi) = \mu^2_\Delta + \frac{1}{2} \l_{H\Delta} \phi^2.
\eea
\eesub
The one-loop correction to the scalar potential induced in presence of $T \neq 0$ reads
\bea
\Delta V_1(\phi,T) &=& \frac{T^4}{2 \pi^2} \bigg[\sum_{a = \text{bosons}} n_a J_B[m^2_a(\phi)/T^2] \nonumber \\
&&
 + \sum_{a = \text{fermions}} n_a J_F[m^2_a(\phi)/T^2] \bigg].
\eea
Here, $J_{B/F}(y) = \int_0^\infty dx~x^2~\text{log}\big[1 \mp e^{-\sqrt{x^2 + y}} \big]$. Further, the infrared effects are included by the daisy remmuation technique~\cite{Weinberg,Dolan,Kirzhnits:1976ts,Carrington:1991hz,Gross,Fendley:1987ef,Kapusta,Arnold:1992rz,Parwani:1991gq}. We adopt the Arnold-Espinosa prescription~\cite{Arnold:1992rz} for the daisy term. That is,
\bea
\Delta V_{\text{ring}}(\phi,T) &=& -\frac{T}{12 \pi} \sum_{a = \text{bosons}} \Big[m_a^3(\phi,T) - m_a^3(\phi) \Big]. 
\eea
The total potential is a sum of the individual contributions,
\bea
V(\phi,T) &=& V_0(\phi) + \Delta V_{\rm CW}(\phi) \nonumber \\
&&
 +  \Delta V_1(\phi,T) + \Delta V_{\text{ring}}(\phi,T).
\eea
We develop a semi-analytic treatment of $V_{\text{total}}(\phi,T)$ in order to gain insight on the $T\neq0$ dynamics. Neglecting the contributions coming from the SM bosons, one finds
\bea
V(\phi,T) \simeq A(T) \phi^2 + C(T)(\phi^2 + K(T))^{\frac{3}{2}} +  B(T) \phi^4. \label{VT_approx}
\eea
A similar approach is followed in~\cite{Quiros:1999jp,Bandyopadhyay:2021ipw}. The various coefficients are computed to be
\besub
\bea
A(T) &=& -\frac{1}{2}\mu^2_H + \frac{1}{32 \pi^2} \bigg[ - 6 y^2_t M^2_t(v)\nonumber \\
&&
 + \frac{3 N \l_{H\Delta}}{2} \Big(m^2_{T_0}(v) + \mu^2_\Delta~\text{log}\frac{T^2}{m^2_{T_0}(v)} \Big) \bigg] \nonumber \\
&&
 + \frac{1}{16}(2 y^2_t + N \l_{H\Delta}) T^2, \\
B(T) &=& \frac{1}{4}\bigg[\l_H + \frac{1}{16\pi^2}\Big(-6 y^2_t ~\text{log}\frac{T^2}{m^2_t(v)} \nonumber \\
&&
 + N\frac{3 \l^2_{H\Delta}}{4}~\text{log}\frac{T^2}{m^2_{T_0}(v)} \Big) \bigg], \\
C(T) &=& - \frac{N}{4 \pi} \bigg(\frac{\l_{H\Delta}}{2}\bigg)^\frac{3}{2}T, \\
K(T) &=& \frac{2 \mu^2_\Delta}{\l_{H\Delta}} \nonumber \\
&&
 + \bigg[\frac{2 \l_{H\Delta} + (3 N + 2)\l_\Delta + 3 g^2_2}{6 \l_{H\Delta}}\bigg] T^2.
\eea
\eesub
Eq.(\ref{VT_approx}) throws light on how the shape of the potential changes with $T$. An extrema at $\phi=0$ is readily identified and it is a minima (maxima) for $A(T) > 0$ ($A(T) < 0$). One thus identifies a threshold temperature $T_0$ determined through $A(T_0)=0$ such that $\phi=0$ is a minima only when $T>T_0$. Another temperature threshold above $T=T_0$ is $T=T_{\text{infl}}$ in which case $V_{\text{total}}(\phi,T)$ develops an inflection point. One solves for $T=T_{\text{infl}}$ through $I^2(T_{\text{infl}}) \equiv 9 C^2(T_{\text{infl}}) - 32 A(T_{\text{infl}}) B(T_{\text{infl}})  + 64 B^2(T_{\text{infl}}) K(T_{\text{infl}}) = 0$. For a temperature range $T_0 < T < T_{\text{infl}}$, the inflection point disappears and a maxima and a minima appear at 
\bea
\phi_{\text{max,min}}(T) &=& 
\bigg[ \frac{1}{32 \big(B(T)\big)^2}
\big[9 C^2(T) - 16 A(T) B(T) \nonumber \\
&&
\pm 3 C(T) I(T) \big] \bigg]^{1/2}
\eea
The temperature range $T_0 < T < T_{\text{infl}}$ is the most interesting from the perspective of first order phase transitions on account of coexisting vacua at $\phi = 0,\phi_{\text{min}}(T)$. One finds a critical temperature $T_c$ in the aforesaid temperature range for which the two vacua are degenerate, i.e., $V(0,T_c)=V(\phi_{\text{min}}(T_c),T_c)$. A strong first order phase transition is identified by $\frac{\phi_{\text{min}}(T_c)}{T_c} > 1$.

\begin{figure*}[!htb]
\centering
\includegraphics[scale=0.40]{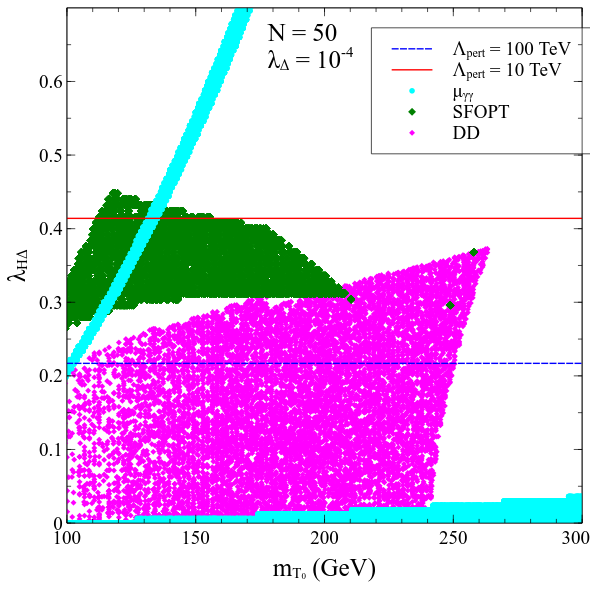}~~~
\includegraphics[scale=0.40]{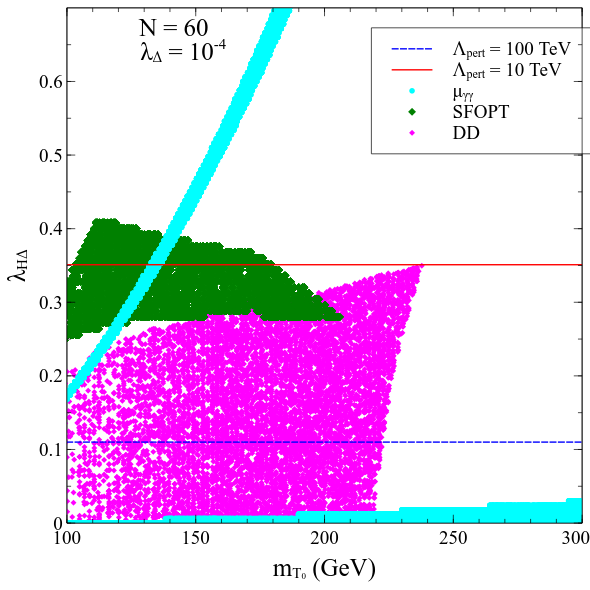} \\
\includegraphics[scale=0.40]{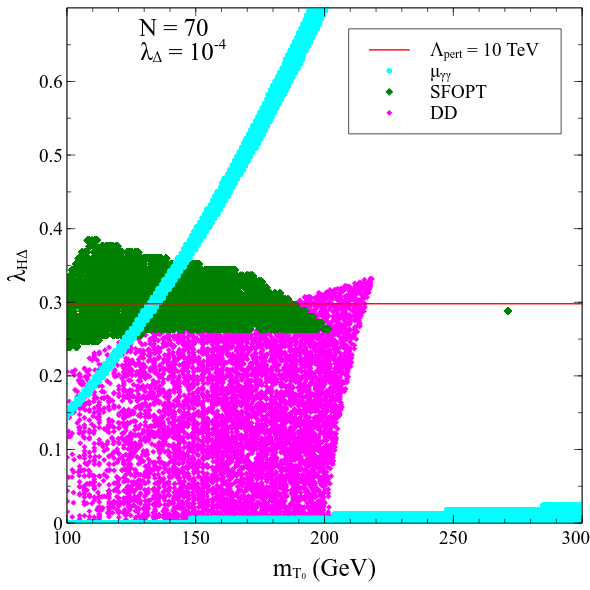}~~~
\includegraphics[scale=0.40]{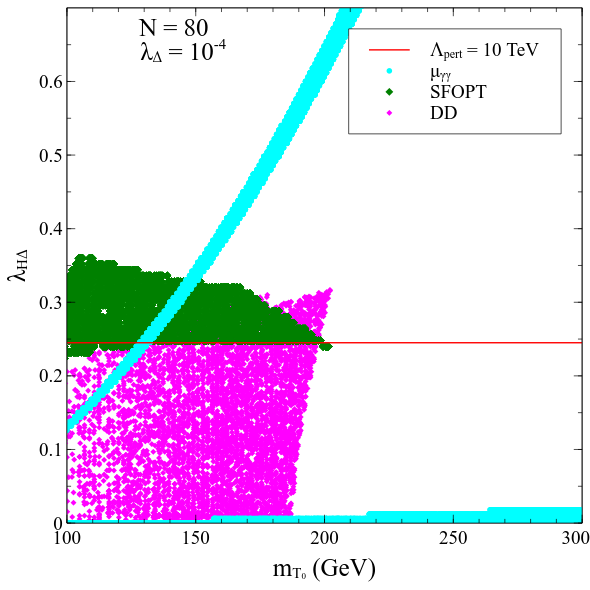} \\
\includegraphics[scale=0.40]{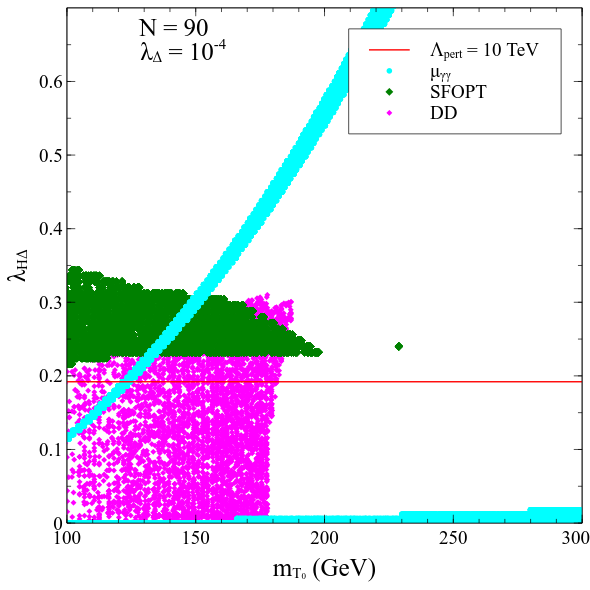}~~~
\includegraphics[scale=0.40]{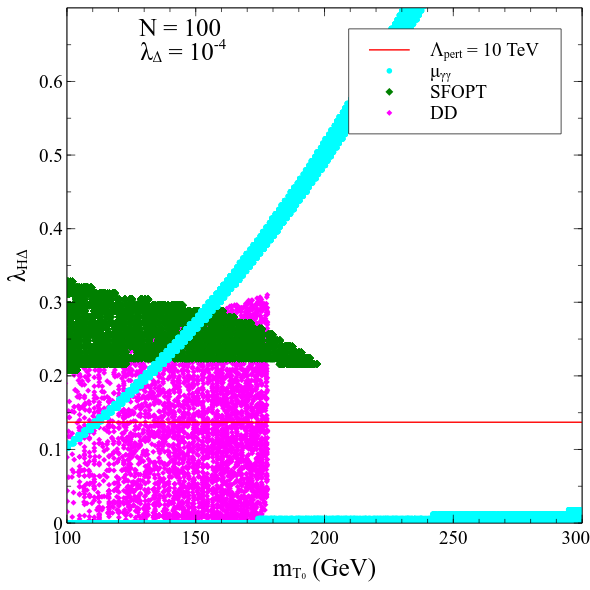} \\
\caption{Allowed parameter regions compatible with $h \to \gamma\gamma$ (cyan), direct detection (magenta) and SFOEWPhT (green) for fixed $N$ and $\l_{H\Delta}$. The horizontal lines denote the scales at which the theory remains perturbative.}
\label{f:mT_lHD}
\end{figure*}

Next, we demonstrate the interplay of the various constraints and SFOEWPhT. The publicly available tool \texttt{PhaseTracer}~\cite{Athron:2020sbe} is used for this part. Fig.\ref{f:mT_lHD} shows regions in the $m_{T_0}-\l_{H\Delta}$ plane that are consistent with $\frac{\phi_{\text{min}}(T_c)}{T_c} > 1$ and the $h \to \gamma\gamma$ and dark matter constraints. The multiplicity $N$ and the parameter $\l_\Delta$ are held fixed. The semi-analytic treatment shows that the lower is $\l_\Delta$, the stronger is the phase transition. We thus choose $\l_\Delta = 10^{-4}$ in the subsequent analysis. In addition, the framework is perturbative up to a given scale in the region below the corresponding horizontal line.

Fig.\ref{f:mT_lHD} shows the results for $N=50,60,70,80,90$ and 100. The diphoton constraint leads to an allowed band in the $m_{T_0}-\l_{H\Delta}$ plane for the $\l_{H\Delta} \gtrsim 0.1$ region. As $N$ increases, the band moves downward in the parameter plane. This trend can be understood from the expression for the $h \to \gamma\gamma$ amplitude in Eq.(\ref{htogaga_np}). The maximum scale for perturbativity, i.e, $\Lambda_{\text{pert}}$, is decided by the choice of $N$ as well as the initial values of $\l_{H\Delta},\l_\Delta$. It is idependent of the triplet mass and this explains the horizontal line. Therefore, the line corresponding to a given $\Lambda_{\text{pert}}$ expectedly moves downward in the parameter plane as $N$ increases. On another part, a SFOEWPhT in a typical Higgs portal setup is known to demand an $\mathcal{O}(1)$ magnitude for the portal coupling. However, this value decreases upon increasing the multiplicity $N$. This is also corroborated by Fig.\ref{f:mT_lHD}. In all, it is seen that there is always an overlap between the parameter spaces compatible with SFOEWPhT and the diphoton constraint. And these overlapping regions are perturbative till scales between 1 TeV and 100 TeV.

However, the direct detection constraint hugely restricts the possibilities. And this is where the present study differs from \cite{Ahriche:2018rao,Bandyopadhyay:2021ipw}. In fact, in intervals of 10 for $N$, the lowest value for the multiplicity for which a SFOEWPhT becomes compatible with $h \to \gamma\gamma$ as well as direct detection is $N=70$. The common parameter patch remians perturbative up to a scale between 10-100 TeV. As $N$ increases, the overlapping patch grows and the maximum scale of perturbativity accordingly gets lowered.
In all, that we can approximately identify $N \simeq$ 70 as the minimum number of hyperchargeless scalar triplets is the major upshot of this analysis. 

Gravitational waves (GW) are inevitably associated
with a strong first order phase transition~\cite{Kamionkowski:1993fg,Witten:1984rs,Hogan:1986qda,Grojean:2006bp}. The three sources that contribute to such a spectrum are, (a) bubble collision~\cite{Kosowsky,Turner,Huber_2008,
Watkins,Marc,Caprini_2008}, (b) sound waves~\cite{Hindmarsh,Leitao:2012tx,Giblin:2013kea,Giblin:2014qia,Hindmarsh_2015}, and, (c) magnetohydrodynamic turbulence~\cite{Chiara,Kahniashvili,Kahniashvili:2008pe,Kahniashvili:2009mf,Caprini:2009yp}. The net GW amplitude thus reads,
\bea
\Omega_{\text{GW}} h^2 &=& \Omega_{\text{coll}} h^2 + \Omega_{\text{sw}} h^2 + \Omega_{\text{tur}} h^2.
\eea
We now list some important parameters that appear in the individual amplitudes. First, the nucleation temperature $T_n$ is determined through $\frac{S_3(T_n)}{T_n} = 140$. Here, $S_3$ refers to the Euclidean action in $d$ = 3 or simply "bounce". The parameter $\beta$ is defined as
\bea
\frac{\beta}{H} &=&\bigg[ T \frac{d}{d T}\Big(\frac{S_3}{T} \Big) \bigg]_{T_n}. \label{beta_expr}
\eea
One must note that value of the Hubble parameter $H$ in Eq.(\ref{beta_expr}) must be computed at $T = T_n$. Next, the energy budget of the phase transition during the bubble nucleation is given by
\bea
\epsilon &=& \Delta V_{\text{tot}}(T_n) - \bigg[T \frac{d \Delta V_{\text{tot}}(T)}{d T}\bigg]_{T_n}.
\eea
In the above, $\Delta V_{\text{tot}}(T) = V_{\text{tot}}(0,T) - V_{\text{tot}}(\phi_{min}(T),T)$ measures the difference in the depths of the potential at the two vacua at a temperature $T$. In addition energy density during nucleantion expressed as 
$\rho_n = \frac{g_*\pi^2}{30}T_n^4$, where $g_*$ denotes the number of degrees of freedom. One subsequently defines
\bea
\alpha &=& \frac{\epsilon}{\rho_n}.
\eea
The quantities $\alpha$ and $\beta$ are of paramount importance in determining the strength of the GW amplitude. The bubble wall velocity $v_b$ used in the above equation is estimated as~\cite{Chao:2017vrq,Dev:2019njv,Paul}
\bea
v_b &=& \frac{1/\sqrt{3} + \sqrt{\a^2 + 2/3\a}}{1 + \a}.
\eea
The GW amplitude of a given source peaks at a given frequency. Such frequencies for the contributions coming from bubble collisions, sound waves and turbulence are are expressed as 
\besub
\bea
f_\text{col} &=& 1.65 \times 10^{-5} 
\Big( \frac{0.62}{1.8 - 0.1 v_b + v^2_b} \Big) \nonumber \\
&&
 \frac{\beta}{H} \Big(\frac{T_n}{100}\Big) \Big(\frac{g_*}{100}\Big)^{\frac{1}{6}}, \\
f_\text{sw} &=& 1.9 \times 10^{-5} 
\Big( \frac{1}{v_b} \Big) \frac{\beta}{H} \Big(\frac{T_n}{100}\Big) \Big(\frac{g_*}{100}\Big)^{\frac{1}{6}}, \\
f_\text{tur} &=& 2.7 \times 10^{-5} 
\Big( \frac{1}{v_b} \Big) \frac{\beta}{H} \Big(\frac{T_n}{100}\Big) \Big(\frac{g_*}{100}\Big)^{\frac{1}{6}}.
\eea
\eesub

%
%

Finally, we express the GW amplitudes from the three sources as a function of frequency $f$ below.
\bea
\Omega_{\text{coll}}(f) &=&  1.67 \times 10^{-5} \Big( \frac{\beta}{H} \Big)^{-2}\Big(\frac{0.11 v^3_b}{0.42 + v^2_b}\Big)
\Big( \frac{\kappa \a}{1 + \a} \Big)^2 \nonumber \\
&&
 \Big(\frac{g_*}{100}\Big)^{-\frac{1}{3}} 
\Big( \frac{3.8 (f/f_{\text{coll}})^{2.8}}{1+2.8(f/f_{\text{coll}})^{3.8}} \Big)
\eea
\bea
\Omega_{\text{sw}}(f) &=&  2.65 \times 10^{-6} \Big( \frac{\beta}{H} \Big)^{-2} v_b
\Big( \frac{\kappa_v \a}{1 + \a} \Big)^2  \nonumber \\
&&
 \Big(\frac{g_*}{100}\Big)^{-\frac{1}{3}} 
\Big(\frac{f}{f_{\text{sw}}} \Big)^3
\Big( \frac{7}{4+3(f/f_{\text{sw}})^{2}} \Big)^2,
\eea
\bea
\Omega_{\text{tur}}(f) &=&  3.35 \times 10^{-4} \Big( \frac{\beta}{H} \Big)^{-2} v_b
\Big( \frac{\epsilon \kappa_v \a}{1 + \a} 
 \Big)^{1.5} \nonumber \\
&&
  \Big(\frac{g_*}{100}\Big)^{-\frac{1}{3}} 
\Big(\frac{f}{f_{\text{tur}}} \Big)^3
\frac{(1 + f/f_{\text{tur}})^{-11/3}}{1+ 8\pi f/h_s}. 
\eea

\begin{figure}
\centering
\includegraphics[scale=0.47]{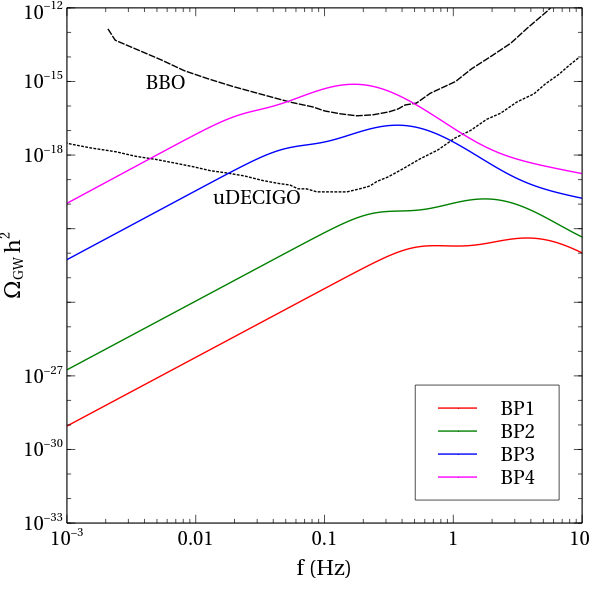}
\caption{GW spectrum of the chosen BPs. The color cosing is explained in the legends. The black lines depict the proposed reach of the detectors BBO and ultimate-DECIGO.}
\label{f:omegaGW}
\end{figure}

We propose below the following benchmark points (BPs) in Table \ref{tab:BP} to test the strength of GW spectrum. We consider $N \geq 70$ while choosing BPs. This is in order to make the SFOEWPhT region compatible with the XENON 1T central values and $h \to \gamma \gamma$ simultaneously, as illustrated in the previous section. And this leads to $\l_{H\Delta} \simeq 0.26$ for all the BPs.
The nucleation temperature $T_n$ and $\frac{\beta}{H}$ are determined using the publicly available tool \texttt{FindBounce}~\cite{Guada:2020xnz}. Fig.\ref{f:omegaGW} displays the total GW amplitudes for the chosen BPs. These BPs are characterised by a steady increase in the phase transition strength as one goes from $N=70$ to $N=100$. The corresponding $\alpha$ values accordingly increase from $\mathcal{O}(0.01)$ to $\mathcal{O}(0.1)$. And finally, $\frac{\beta}{H}$ is seen to decrease by almost two orders. The GW amplitudes increase with increasing $\alpha$ and lowering $\frac{\beta}{H}$. Accordingly, we see in Fig.\ref{f:omegaGW} that BP1 and BP2 remain beyond the reach of the proposed detectors. On the other hand BP3 and BP4 respectively are within the reach of u-DECIGO and BBO respectively.

\begin{table*}[t!]
\centering
\begin{tabular}{ |c|c|c|c|c|c|c|c|c|c| } 
\hline
 & $N$ & $m_{T_0}$ & $\l_{H\Delta}$ & $\l_\Delta$ & $T_c$ & $\frac{\phi_c}{T_c}$ & $T_n$ & $\a$ & $\frac{\beta}{H}$ \\ 
\hline \hline 
BP1 & 70 & 127.531 GeV  & 0.264752 & $10^{-4}$  & 106.698 GeV  & 1.69719 & 106.423 GeV & 0.0379803 & 112744.  \\
BP2 & 80 & 134.717 GeV & 0.257330 & $10^{-4}$ & 113.801 GeV & 2.21102 & 113.108 GeV & 0.0600576 & 50244.3 \\
BP3 & 90 & 141.867 GeV & 0.265049 & $10^{-4}$ & 123.315 GeV & 3.57403 & 119.376 GeV & 0.140405 & 10481.3 \\
BP4 & 100 & 150.448 GeV & 0.271032 & $10^{-4}$ & 140.876 GeV & 4.73334 & 129.447 GeV & 0.219615 & 4562.59 \\ \hline
\end{tabular}
\caption{Benchmark points chosen to study the GW spectrum.}
\label{tab:BP}
\end{table*}

\section{Conclusions and future outlook}\label{conclu}

The $Y=0$ triplet is one of the most minimal extensions of the SM furnishing a scalar DM candidate. In this study, we have generalised such a framework by postulating $N$ such triplets governed by a global $O(N)$ symmetry. This global invariance ensures that the framework is describable in terms of the same Lagrangian parameters as the $N=1$ case, that is, the triplet mass $m_{T_0}$, the Higgs-triplet portal coupling $\l_{H\Delta}$, and, the triplet self couping 
$\l_\Delta$. The multiplicity $N$ is an a priori unknown integer of this framework. It follows that several quantities of interest are sensitive to the value chosen for $N$ in such a scenario.

In this study, we probe SFOEWPhT in the $O(N)$-symmetric scalar triplet scenario. While multi-step phase transitions are permissible in principle, we look for a single-step one for minimality in the direction of the SM Higgs doublet. In the process, we also discuss how this scenario confronts the crucial constraints coming from $h \to \gamma \gamma$ signal strength, DM direct detection and an absence of Landau poles up to a given cutoff. For a given $N$, we identify the area in the $m_{T_0}-\l_{H\Delta}$ plane compatible with the single step SFOEWPhT and the aforementioned constraints. This analysis reveals $N \gtrsim 70$ for the SFOEWPhT to be compatible with the constraints. The direct detection constraints turn out to be hugely restricting in this scenario given the economy of parameters in this setup. We also estimate the strength of the GW signals associated with the SFOEWPhT parameter regions. We find that for GW signals to be observable in the proposed space-based detectors ultimate-DECIGO and BBO, one must have $N \gtrsim 90$. Therefore, the present analysis also zeroes-in on the minimum multiplicity of the triplets required to produce observable GW signatures.

\section{Acknowledgements}

NC acknowledges support from DST,
India, under grant number IFA19-PH237 (INSPIRE Faculty Award). NC also thanks Peter Athron for useful correspondence on \texttt{PhaseTracer}.

	\bibliographystyle{JHEP}
	\bibliography{ref}

\end{document}